\documentstyle[12pt]{article}
\title{MIXED COLD-HOT DARK MATTER MODEL WITH FALLING AND QUASI-FLAT
INITIAL PERTURBATION SPECTRA}

\author {Dmitri Yu. Pogosyan\thanks{On leave of absence from Tartu
Astrophysical Observatory, T\~{o}ravere, EE2444, Estonia}\\
CITA, Univ. of Toronto, Toronto ONT M5S 1A7, Canada \\
pogosyan@cita.utoronto.ca \\
\and Alexei A. Starobinsky\\
Yukawa Institute for Theoretical Physics,\\
Kyoto University, Uji 611, Japan \\ and \\
Landau Institute for Theoretical Physics,\\
Kosygina St. 2, Moscow 117334, Russia\\
alstar@yisun1.yukawa.kyoto-u.ac.jp}

\begin{document}
\maketitle
\clearpage
\begin{abstract}
The mixed cold-hot dark matter cosmological model (CHDM)
with $\Omega_{tot}=1$ and a
falling power-law initial spectrum of Gaussian adiabatic
perturbations ($n>1$) is tested using recent obserbational data.
It is shown that its fit to the data becomes worse with the growth
of $n-1$, and may be considered as unreasonable for $n>1.1$
for all possible values of the Hubble constant. Thus,
the CHDM model with a
falling initial spectrum is worse than the same model with the
approximately flat $(|n-1|<0.1)$ spectrum.
On the other hand, the CHDM model provides a rather good fit to
the data if $n$ lies in the range $(0.9-1.0)$, the Hubble
constant $H_0 < 60$ km/s/Mpc ($H_0 < 55$ for $n=1$) and
the neutrino energy density $\Omega_{\nu}< 0.25$.
So, the CHDM model provides the best possibility for the
realization of the simplest variants of the inflationary scenario
having the effective slope $n\approx (0.95-0.97)$ between galaxy and
horizon scales, including
a modest contribution of primordial gravitational wave background to
large-angle $\Delta T/T$ fluctuations of the cosmic microwave
background (resulting in the
increase of their total {\it rms} amplitude by $(5-10)\%$) expected in
some variants. A classification of cosmological models according
to the number of fundamental parameters used to fit observational
data is presented, too.

\end {abstract}

\section{Introduction}

It remains an ambitious goal of the inflationary scenario, as well as of
any other fundamental cosmological theory of the early Universe, to
explain all observed structure of the present-day Universe using a
minimal number of
additional microphysical ``fundamental'' constants, apart from those
already khown from the particle physics. Of course, we don't know how
many parameters is really needed to describe the whole Universe, so
one can't say {\it apriori} that a theory having more parameters is
worse than a theory with a less number of them. However, following the
the {\it Occam's razor} principle, classification of different
cosmological models according to the number of additional
phenomenological parameters
used in them gives us a natural {\it logical} sequence of their
consideration and comparison with observational data.
We call these parameters fundamental, if they appear in basic
equations (as in the inflationary scenario), not in initial conditions
or other assumptions.

As is well known, a power spectrum of perturbations producing the
observed structure of the Universe is a product of an initial
(primordial) spectrum formed in the early Universe
and a transfer function $C^{2}(k)$ which depends on
the type of dark matter at present (e.g., on masses and concentrations
of neutrinos).  From the inflationary scenario point of view, the
initial spectrum is completely determined by a phenomenological
Lagrangian of an effective scalar field (or fields) -- inflaton(s) --
at the de Sitter (inflationary) stage in the very early Universe. Then
parameters determining both the initial spectrum and the present dark
matter content are fundamental and should be considered and counted
on equal footing.
If such a classification is applied to inflationary models (see, e.g.,
Starobinsky 1993), then a model of the first level having only one
fundamental parameter -- an amplitude of perturbations --
is the CDM model with the approximately flat
(Harrison-Zeldovich, or $n\approx 1$) spectrum of initial adiabatic
perturbations. Because of theoretical considerations and observational
uncertainties, it is better to include ``weakly-tilted'' models with
$|n-1|\le 0.1$ into this class, too. Hereafter this model will be
referred as the Standard Cold Dark Model (SCDM). There exists another
model belonging to this level: the CDM model with the approximately
flat spectrum of isocurvature fluctuations ($n\approx -3$). But that
model has been known to be excluded by observations long ago,
because it produces excessive large-scale $\Delta T/T$ fluctuations.
Strictly speaking, SCDM has one more parameter which defines an
amplitude of the approximately flat spectrum of primordial gravitational
waves and which is directly connected to the Hubble parameter
$H\equiv \dot a/a$ at the de Sitter stage (Starobinsky 1979),
$a(t)$  being the scale factor of the Friedmann-Robertson-Walker
cosmological model. However, this is a rather small effect which may
be seen in a slight increase of large-scale $\Delta T/ T$
fluctuations only (apart, of course, from a remote possibility of
direct detection of this relic gravitational background), see
the discussion section below. So, it is
better to consider parameters connected with the gravitational wave
background separately.

At present, it is clear already that SCDM predictions, though being
not far from  observational data (that is remarkable for a
such a simple model with only one free parameter),
still definitely do not agree with all of them.
Namely, if the free parameter is chosen to fit the data on scales
$(100-1000)h_{50}^{-1}$ Mpc, discrepancy of about twice in perturbation
amplitude arises on scales $(1-10)h_{50}^{-1}$ Mpc,
and vice versa ($h_{50}=H_0/50$, where $H_0$ is the Hubble constant
in km/s/Mpc). Thus, models of
the next (second) level having one more additional constant have
to be considered. Among these models, the best is certainly the mixed
cold+hot dark matter model (CHDM), Shafi \& Stecker 1984,
for recent analysis see Pogosyan \& Starobinsky 1993
(hereafter PS), Liddle \& Lyth 1993, Klypin et al. 1993
and references therein. In this model, the hot component is
assumed to be the most massive of 3 neutrino species (presumably,
$\tau$-neutrino) with the standard concentration following from
the textbook Big Bang theory. Then the only new fundamental parameter
is the neutrino mass $m_{\nu}$ (masses of the other two types of
neutrinos are supposed to be much less and, therefore, unimportant for
cosmology). If, on the contrary, masses of two neutrino types are
assumed to be
comparable or even equal, the resulting model will belong to the third
level, until the mass ratio will be either confirmed in laboratory
experiments, or
theoretically derived from some underlying theory (we shall
return to the discussion of this case at the end of the
paper). $m_{\nu}$ is related to the energy density of the
hot component (in terms of the critical one) $\Omega_{\nu}$ by
\begin{equation}
m_{\nu}=23.3\Omega_{\nu}h_{50}^2~eV
\label{mass}
\end{equation}
for $T_{\gamma}=2.735$K. The CDM model with the cosmological constant
seems to be on the second place by a number of difficulties (and still
marginally admissible), and the two tilted CDM models with a power-law
initial spectrum of adiabatic perturbations (with $n<1$) and
isocurvature ones (with $n>-3$ to avoid excessive large-scale
$\Delta T/T$ fluctuations) are marginally or completely excluded.

Still the CHDM model with $n\approx 1$ is not without difficulties.
The main of them is connected with later galaxy and quasar formation
in this model as compared to the SCDM model. As a result, only a small
region in the $H_0-\Omega_{\nu}$ plain remains permitted (PS, see also
a more pessimistic view in Cen \& Ostriker 1994).
Recently, this difficulty exacerbated
due to the problem of producing sufficient number of damped
$Ly-\alpha$ systems (Subramanian \& Padmanabhan 1994, Mo \&
Miralda-Escude 1994, Kauffmann \& Charlot 1994). However, latest
analysis based on $N$-body simulations suggests that the latter problem
may be solved if $\Omega_{\nu}$ is taken smaller than it was supposed
before:
$\Omega_{\nu}\le 0.25$ (Klypin et al. 1994) or even $\Omega_{\nu}\le
0.2$ (Ma \& Bertschinger 1994) for $h_{50}=1$, in complete agreement
with the restrictions on the model following from quasar and galaxy
formation which were earlier obtained in PS using linear theory.

Therefore, it is desirable to have more power on small scales in the
CHDM model. This was one motivation for us to consider a CHDM
model belonging to the next (third) level, i.e., having one more
parameter. We assume the standard neutrino concentration and use one
of three adjustable parameters as the neutrino mass as earlier.
Then we are left with two parameters to characterize an
initial perturbation spectrum. The most natural possibility is to
assume a power-law spectrum of adiabatic
perturbations with $n\not=1$, then one of the parameters gives a
rms amplitude of perturbations at some scale (say, at the present
horizon scale), while the other defines the slope. For the reason
stated above, we consider the case $n>1$ in this paper (with some
results relevant to the approximately flat case $|n-1|<0.1$, too).
The case $n<1$ for the CHDM model has been
already considered in detail in Liddle \& Lyth 1993 and briefly
mentioned in PS, with the conclusion that the
``really'' tilted case with $n<0.9$ is excluded, but typical
chaotic inflationary spectra with $n\approx 0.95 - 0.97$ (which we
count as approximately flat ones) are possible.

The other motivation to consider such a model is that the best fit
to the COBE data is given by $n$ slightly larger than $1$: $n\approx
1.2$ (Bennett et al. 1994, G\'orski et al. 1994), although $n=1$ lies
inside $1\sigma$ error bars (and, as advocated by G\'orski et al.
1994, becomes even the best fit if the quadrupole is completely
excluded). $n>1$ is also needed to account for the
results of the Tenerife experiment (Hancock et al. 1994). Of course,
$n$ significantly larger than $1$ produces too much power at small
scales and, thus, may be rejected for different reasons. E.g., if
the perturbation spectrum is assumed to maintain its power-law form
up to very small scales where metric perturbations (gravitational
potential) become comparable to unity, then the upper limit
$n< 1.4$ follows from the consideration of production of primordial
black holes (Carr, Gilbert \& Lidsey 1994).
Even if a falling power-law spectrum
is assumed over a much smaller scale range of a few orders of magnitude,
another upper limit $n<1.54$ follows from the absence of spectral
distorsions of the cosmic microwave background (Hu, Scott \& Silk
1994). For these reasons, we investigate the range $n\le 1.3$ only.

On the other hand, in spite of the advent of the Hubble telescope,
there is still no general agreement about the value of the Hubble
constant $H_0$. One methods produce the value of $H_0$ around $50$
km/s/Mpc (Branch \& Miller 1993, Sandage et al. 1994), others
lead to a significantly larger value $(70-80)$ km/s/Mpc
(Schmidt et al. 1994),
while the accuracy of new methods based, e.g., on the
Sunyaev-Zeldovich effect in rich clusters of galaxies,
is still not enough to discriminate
between these two cases (Birkinshaw \& Hughes 1994). That is why we
have to investigate a dependence of the model predictions on
$H_0$ additionally. We assume  $H_0$ in the range $(40-80)$ km/s/Mpc.
Of course, $H_0$ is neither a new, nor a fundamental parameter,
so it should not be counted in our classification scheme.

Finally, to justify the use of the notion of fundamental
parameters, we have to present at least one inflationary model
producing a falling power-law spectrum over some range of scales.
The simplest way is to
take the well-known case of a test massive scalar field in the de
Sitter background, that is equivalent to assuming the inflaton
potential to be of the form $V(\phi)=V_0+{m^2\phi^2\over 2}$ with
$m^2\sim H_1^2\equiv {8\pi GV_0\over 3}\ll M_P^2$ and
the field $\phi$ being in the regime $|\phi |\ll M_P$ ($M_P
\equiv G^{-1/2}$ is the Planck mass, and $\hbar =c =1$ is used
in this paragraph). The inflationary stage in this model ends
when $\phi$ drops to a very small value $\phi_f \ll M_P$ either
as in the new inflationary model, or due to a second-order phase
transition destroying $V_0$ (Linde 1994). The exact expression
for the slope is $n-1=3-\sqrt{9-4{m^2\over H_1^2}}$. The slope $n=1.1$
corresponds to $m=0.38H_1$ (of course, then it is possible to use
an approximate form of this formula: $n\approx 1+{2m^2\over 3H_1^2}$).
Therefore, we may take $m_{\nu},~{H_1\over M_P}$ and ${m\over H_1}$
as the three new fundamental parameters of the cosmological scenario
considered.

\section{CHDM Linear Perturbation Spectrum}

Following our previous notations for the case of the flat initial
spectrum, we write the linear power spectrum
of perturbations of the gravitational potential $\Phi$ in the form
\begin{equation}
P_{\Phi}(k,t) = {9A^2\over 200\pi^2k^3}
{\left( {2 c k \over H_0}\right) }^{n-1}
\cdot C^2_{\mathrm CHDM} (k,t)~,
\end{equation}
specifying an initial spectrum produced at an early (inflationary)
stage of the evolution of the Universe to be the power-law one:
$ P_{\Phi ,\mathrm in}={9A^2 \over 200\pi^2k^3}{\left( {2 c k \over H_0}
\right) }^{n-1} $.
The transfer function $ C_{\mathrm CHDM} $  was determined in our previous
paper (PS) numerically by solving a system of the Einstein-Vlasov
equations for the evolution of adiabatic perturbations with
a neutrino component treated kinetically and a CDM component as dust.

The resulting $ C_{\mathrm CHDM} $ depends on present values of
the neutrino fractional density
$ \Omega_{\nu} $, the Hubble constant $H_0 = 50 h_{50}$ km/s/Mpc
and the CMB temperature $ T_{\gamma} = 2.735 t_{\gamma}$ K
as parameters, as well as on time $t$ (redshift $z$).
If we define $C_{\mathrm CHDM}(k,\Omega_{\nu},z)=C_{\mathrm CDM}(k)
\cdot D(k,\Omega_{\nu},z) $,
our numerical calculations can be described by the
following fitting formula
\begin{equation}
D(k,\Omega_{\nu},z) = {\left( {{ 1+ (Ak)^2 +
{(1-\Omega_{\nu})}^{1 \over \beta} (a_{\mathrm eq} / a_0) (1+z) (Bk)^4 }
\over {1 + (Bk)^2 - (Bk)^3 + (Bk)^4}} \right) }^{\beta},
\label{transf}
\end{equation}
with $ \beta = {{5-\sqrt{25-24\Omega_{\nu}}} \over 4} ~.$
This formula satisfies asymptotic regimes for the transfer
function discussed in PS. Here $a_0/a_{\mathrm eq} = (1.681~
\Omega_{\gamma})^{-1}$ is the expansion factor of the Universe from
the matter-radiation equality moment $ t_{\mathrm eq} $ until the present
time $ t_0 $.
The scale $ B  = R_{nr} \cdot (1+\Omega_{\nu 0}) /(\Omega_{\nu} +
\Omega_{\nu 0}) $,
where $ R_{nr} = 10.80 \ h_{50}^{-2} t_{\gamma}^2$ Mpc,
is the neutrino nonrelativization scale for $ \Omega_{\nu} = 1$.
For the critical value $ \Omega_{\nu} =\Omega_{\nu 0} = 0.1435 $,
neutrinos become nonrelativistic at $t=t_{eq}$.
The best fit of the only scale remained gives
$ A=R_* \left( 1+ 10.912 \Omega_{\nu} \right)
\sqrt{\Omega_{\nu}(1-0.9465 \Omega_{\nu})} \, /
\left( 1+{(9.259 \Omega_{\nu})}^2 \right)$
with $ R_*=69.52 \ h_{50}^{-2} t_{\gamma}^2$ Mpc. The deviation of
this fit from the calculated $ D(k) $ is better than 2\% in the
region $ k < 5 h_{50}^2$ Mpc$^{-1}$ for the present moment $ z= 0 $
and $ \Omega_{\nu} \le 0.7 $, while staying within 5\% for $ z < 30 $.
We used the Bardeen et al. (1986) expression for $C_{\mathrm CDM}(k) $
in actual fitting.

If the present CMB temperature is fixed (we use
$T_{\gamma} = 2.735$ K), the model involved
has four parameters, namely, $A$, $n$, $\Omega_{\nu}$ and
$H_0$. If the adiabatic mode is solely responsible for the
observed large-angle CMB fluctuations, we have
\begin{equation}
\langle\left( {\Delta T \over T }\right)_{lm}^2 \rangle =
{A^2 \over 100}{ \Gamma (3-n) \over 2^{3-n} \Gamma ^2 (2-{n\over 2})}
{ \Gamma (l+{n-1\over 2}) \over \Gamma (l+2-{n-1\over 2}) }
\label{dt}
\end{equation}
(Bond \& Efstathiou 1987). Note also a simple form of the corresponding
large-angle correlation function (with the dipole
components excluded) which we did not notice in literature:
\begin{eqnarray}
\xi_T(\vartheta) &\equiv& \langle {\Delta T\over T}(0),~{\Delta T
\over T}(\vartheta)\rangle \nonumber \\
& =&{A^2\over 200\pi^2}{\Gamma(n)\cos {\pi (n-1)
\over 2}\over 2^{n-1}(n-1)(2-n)}\left[ \left(\sin {\vartheta
\over 2}\right)^{1-n}-{2\over 3-n}\left(1+{3(n-1)\over (5-n)}\cos
\vartheta \right) \right]  \nonumber \\
&=& a_2^2~{(3-n)(5-n)(7-n)\over 10 (n^2-1)}\left[ \left(\sin
{\vartheta \over 2}\right)^{1-n}-{2\over 3-n}\left(1+{3(n-1)\over
(5-n)}\cos \vartheta \right) \right]    \label{cor}
\end{eqnarray}
where $a_2^2\equiv {5\over 4\pi}\langle\left( {\Delta T \over T}
\right)_{2m}^2 \rangle$.  This expression is valid for both $n>1$
and $-1<n<1$. The known
correlation function for the $n=1$ case (Starobinsky 1983) follows
from here by limiting transition. If the tensor mode (gravitational waves)
is responsible for a part of observed $ \Delta T / T $
fluctuations, the value $ A $ derived from (\ref{dt}) for given
$\langle{\left( \Delta T \over T \right)}_{lm}^2 \rangle $
serves as an upper limit on the amplitude of the adiabatic mode.
However, we don't expect a noticeable tensor contribution for  $n>1$,
unless a new parameter is introduced into the model that would
shift it into the next, fourth level.

\section{Confrontation with Observational Tests}

Our way of comparison of the model with observational tests closely
follows our previous paper (PS).
However, here we confine our consideration to the following
tests shown to be the most restrictive for the CHDM model in PS.
\begin{enumerate}
\item The COBE measurement of large-angle $ \Delta T / T $
fluctuations.
\item Value of the total {\it rms} mass fluctuation $\sigma_8$
at $R=16h{^{-1}_{50}}{\mathrm Mpc}$ (we use the index ``$8$'', not
``$16$'', to be in accordance with standard notation caused by the habit
of measuring $H_0$ in units of $100$ km/s/Mpc).
\item The Stromlo-APM counts in cells (Loveday et al. 1992).
\item Density of quasars (Efstathiou \& Rees 1988, Haehnelt 1993).
\item Large-scale peculiar velocities following from the POTENT
reconstruction (Bertschinger et al. 1990, Dekel 1993).
\end{enumerate}
Other tests included in the extended list of PS do not lead to
additional limitations on the model.

We adopted the following numerical values for the observational
data considered.

The COBE result for the total {\it rms} value of the $l=4$ multipole
$(\Delta T / T) _4 = (12.8 \pm 2.3)$ $\mu$K$/T_{\gamma}$
(Wright et al. 1993) is used
to put limits on the amplitude $A$, because it seems to be the most
spectrum independent. For $n=1$, this corresponds to the amplitude
$ A = (4.38 \pm 0.79)\times 10^{-4} $ and the total quadrupole value
$Q_{rms-PS} = (17.4 \pm 3.1)$ $\mu$K
which are somewhat higher than those used in PS.

The total {\it rms} mass fluctuation $\sigma_8 \equiv \left({\delta M
\over M}(16h{^{-1}_{50}}{\mathrm Mpc})\right)_
{tot}$ is calculated for the {\it lower} limit of the COBE amplitude
$ (\Delta T / T)_4 = 10.5$ $\mu$K$/T_{\gamma}$ (note that the
inverse quantity $\sigma_8^{-1}$
is usually understood as the biasing parameter $b$ for optical galaxies).
Based on cluster
abundance data (White, Efstathiou \& Frenk 1993), we consider
$\sigma_8 <0.67$ as a conservative upper limit on $\sigma_8$.

The Stromplo-APM counts-in-cells represent nine data points
for the mass variance in cubic cells $\sigma^2 (l)$ over the range
$l = (20 - 150) \, h_{50}^{-1}$ Mpc. Fixing normalization by the
condition $\sigma^2 (l) =1$ for $l=25 \, h^{-1}_{50}$ Mpc which
agrees with the data, we eliminate a constant redshift correction
(Kaiser 1987) and become able to compare directly the shape
of the power spectrum with theoretical predictions for these
scales. To find the best fit to the data, we formally applied
the $\chi^2$ test with $N=9-2=7$ degrees of freedom for a fixed $n$.

The standard model of quasar formation assumes that they arise as a
result of formation of massive black holes in nuclei of galaxies with
total masses $(10^{11} - 10^{12}) \, {\mathrm M}_{\odot} $. Recent
estimates of the fraction of mass in bound objects which can serve
as quasar hosts at $z=4$ give
$ f( \ge 10^{11} \, {\mathrm M}_{\odot} ) \ge 10^{-4} $ (Haehnelt 1993).
In this paper, Haehnelt also presented the estimate for a fraction of
mass in large galaxies: $ f( \ge 10^{12} \, {\mathrm M}_{\odot} ) \ge
10^{-5} $ at the same redshift $z=4$. We use the
simple Press-Schechter formalism to connect the mass fraction $f(\ge M)$
to a linear mass fluctuation $\sigma (M,z) $ on a scale $M$ at a
redshift $ z $:
\begin{equation}
f(\ge M) = 1 - erfc \left( { \delta_c \over \sqrt{2} \sigma (M,z) }
\right)~. \label{ps}
\end{equation}
Then we get the limitation
\begin{equation}
\left[(1+z)~\sigma (M,z) \right]_{z=4}\ge \alpha \delta _c
\label{s11}
\end{equation}
where $\alpha = 1.285~(1.132) $ for $M=10^{11}~(10^{12})~
{\mathrm M}_{\odot}$. The left-hand side of Eq. (\ref{s11}) depends on
$z$ only due to the $z$-dependence of the transfer function (\ref{transf}).
There is no consensus on the threshold value $\delta _c $ to be used.
The standard one for top-hat fluctuations is $ \delta_c = 1.69 $, while
Klypin et al. (1994) advocate for $\delta_c = 1.4 $ as the best fit
to a mass distribution in the CHDM model at high $z$. In the latter
case, however, the Gaussian filtering was used to calculate $\sigma (M)$
for a given mass scale.
Let us note that the choice $M=10^{12} \, {\mathrm M}_{\odot} $
provides a tighter limitation than $M=10^{11} \, {\mathrm M}_{\odot} $.
Therefore, we adopt (\ref{s11}) with $\delta _c = 1.4$ (but with the
Gaussian
filtering) as a conservative restriction, while the limit on $ 10^{12}\,
{\mathrm M}_{\odot} $ objects with the more standard $\delta _c = 1.69$
for top-hat fluctuations as, may be, more realistic one. Also, we found
useful to use an amplitude-independent combination of the $\sigma_8$
limitation $\sigma_8<\sigma_*$ and the number of quasars test in the form
\begin{equation}
\left[(1+z)~\sigma (M,z) \right]_{z=4}\sigma _8^{-1} \ge \alpha \delta _c
\sigma_*^{-1}~.
\label{qb}
\end{equation}

 From the POTENT data, we use two values of large-scale
bulk velocities  $ v(80 h_{50}^{-1} \, {\mathrm Mpc}) = (405 \pm 60)$ km/s
and $v(120 h_{50}^{-1} \, {\mathrm Mpc}) = (340 \pm 50)$ km/s (Dekel 1993).
These values are in agreement with results
obtained by other groups, see, e.g., Courteau et al. 1993 where the
values $v(80 h_{50}^{-1} \, {\mathrm Mpc}) = (385 \pm 38)$ km/s and
$v(120 h_{50}^{-1} \, {\mathrm Mpc}) = (360 \pm 40)$ km/s for our bulk
motion are presented. Note, however, that all these values refer to
bulk flow velocities in our vicinity and, thus, may differ from
their {\it rms} values calculated for the whole Universe (``cosmic
variance'').

In Fig. 1, we display restrictions in the $H_0 - \Omega _{\nu}$ plane
for several values of the slope of initial spectrum $n=0.85,0.95,1.1,1.2$
based on a combination of tests, namely the counts-in-cells $\sigma^2(l)$
values and the combined quasar density -- $\sigma_8$ condition
(\ref{qb}). The case $n=1.0$ was extensively discussed in PS (see
Fig. 6 therein). The main conclusion made is that CHDM
model parameters are restricted to the narrow range of
a low Hubble constant $ H_0 < 60 $ km/s/Mpc and
the neutrino fraction $\Omega_{\nu} = 0.17-0.28 $ for $H_0 = 50 $.
The first of these limits reflects a problem with unavoidable
high mass fluctuations at the $16 h_{50}^{-1}$ Mpc scale, as well as
a wrong shape of the perturbation spectrum over the $l=(20-150)
h_{50}^{-1}$ Mpc interval if the Hubble
constant is high. The upper bound on $\Omega_{\nu}$ comes from
the combined condition (\ref{qb}) that implies that the slope
of the spectrum in the scale range $(0.7 h_{50}^{-2/3} -
16 h_{50}^{-1})$ Mpc cannot be too steep. The lower
bound on $\Omega_{\nu}$ arises both by matching of the spectrum slope
to the counts-in-cells $\sigma ^2 (l)$ values (that shows more relative
power on large scales than in the SCDM model), as well as from the
result that $\sigma_8$ is too high for low-$\Omega_{\nu}$
models with the COBE normalization.

Fig. 1 shows how these results are affected by allowing an initial
power-law spectrum to be non-flat: $n \ne 1$.
Two main conclusions can be drawn. The amplitude independent tests
remain in essentialy the same mutual relation favouring somewhat
higher values of $\Omega_{\nu}$ for larger $n$ as expected . For
illustration, one may follow the point of intersection
of the $\chi ^2 = 2 $ contour with the limiting line for
$10^{11} \, {\mathrm M}_{\odot} $ objects. This point moves from
$H_0 = 62,~\Omega_{\nu} = 0.3$ for $n=0.85$ to $H_0 = 52,~
\Omega_{\nu} = 0.38$ for $n=1.2$. On the other hand,
restrictions from the $\sigma_8$ test become
much tighter as the slope of initial spectrum increases. For $n=1.2$,
all models with $H_0 \ge 50 $ are antibiased: $ b < 1 $. Thus,
models with large $n$ have too large mass fluctuations at the $16
h_{50}^{-1}$ Mpc scale that cannot be compensated by increasing
the neutrino fraction $\Omega_{\nu}$.

We present another look at these results by plotting
the $n - \Omega_{\nu} $ cross-section of the parameter space
for the same three tests in Fig. 2 . One can see that, for
low values of the Hubble constant $H_0=40,~50 $, there is a
significant degree of degeneracy between $n$ and $\Omega_{\nu}$,
as far as the tests on the spectrum shape are considered,
which do not disallow any value of $n$. Note, however, that if
we adopt the more stringent quasar abundance condition
$ f(\ge 10^{12} \, {\mathrm M}_{\odot}) \ge 10^{-5} $ at $z=4 $,
then we get an {\it upper} limit on $\Omega_{\nu}$ for $H_0=50$
roughly coinciding with the {\it lower} boundary
of the best $\chi ^2 =2$ region for the counts-in-cells fit. So
both tests are only in a marginal agreement for parameters
close to the line $3 \Omega_{\nu} = n-1/4 $.

As we increase $H_0$, the fit to the counts-in-cells values begins
to fail for large spectrum indexes, so that the best $\chi ^2 = 2$
contour sets the limit $ n < 1.1 $ for $H_0 = 60 $,
while for $H_0 = 70 $, this limit follows even from the more
conservative condition $\chi ^2 \le 7 $. Moreover, the best-fit
contours tend to select higher values of the neutrino fraction than is
allowed by the quasar test. In this way, for $H_0 \ge 55 $, no model
can satisfy
the condition (\ref{s11}) for $M=10^{12} \, {\mathrm M}_{\odot}$ and have
$\chi^2 < 2 $ (for $H_0 > 70 $, even $\chi^2 < 7 $) simultaneously.
If the less stringent quasar test for
$M=10^{11} \, {\mathrm M}_{\odot}$ is used, it becomes possible to achieve
the best fit for the counts-in-cells data for $H_0=60$ if $n < 0.9$,
but not for the Hubble constant as high as $H_0=70$. On this basis, we
conclude once again that the CHDM model is incompatible with high
values of the Hubble constant $H_0 > 60 $ even for
$n\not =1$ initial spectra.

Although tests on the spectrum shape do set a tight limitation
on the initial slope for high $H_0$, too, the strongest
argument against high $n$ comes directly from the $\sigma_8$ condition.
To achieve $\sigma_8<0.67$ for $H_0=50$, one must restrict the model
to $n \le 1$, while for $H_0=60$, even $\sigma_8<1 $ leads to $n < 1.05$.

All the tests on the CHDM model become more in agreement with each
other for
$H_0 = 40 $. However, we are not sure that such low values of the Hubble
constant are possible. Therefore, on the basis of Figs. 1,~2, we consider
a small region of the parameter space in the vicinity of
$H_0=50,~\Omega_{\nu} = 0.23 ,~n=0.95$ as the best parameter set
for the model. It corresponds to the neutrino mass $m_{\nu}\approx 5$ eV.

The Stromplo-APM counts-in-cells $\sigma ^2 (l)$ are given in redshift
space.
Although we excluded any constant redshift correction using these data
as a test on the slope of the initial spectrum, it may be asked how
the results would change if the redshift correction depended on
scale significantly. Here we note that the counts-in-cells
test in the form we used it serves primarily to indicate the
necessity for relatively high $\Omega_{\nu}$.
Probably, the redshift correction can only increase with scale.
Then underlying real space mass fluctuations depend less steeply
on scale and are better fitted by lower $\Omega_{\nu}$ models than the
straightforward use of $\sigma ^2 (l)$ predicts.

To support the conclusions derived from the counts-in-cells test,
we present $\chi ^2 $ contours of the direct fit of theoretical power
spectra to a power spectrum reconstructed from the galaxy angular
correlation function $w(\theta )$ (Baugh \& Efstathiou 1993, 1994)
in Fig. 3. These data give the power spectrum directly in real space.
We used their values for $P(k)$ in the range $k=(0.07 - 0.025) h_{50}$
Mpc$^{-1}$ where errors are not as large as on larger scales and
nonlinear corrections are not yet as
important as on smaller scales. In fact, we have only 4 data points
in the considered range which are fitted with the 3-parametric model
(for fixed $n$ or $H_0$).
In Fig. 3, we compare $\chi ^2$ contours of this one
and the counts-in-cells tests in the $H_0 - \Omega _{\nu}$ plane
for $n=0.95$ and in the $n - \Omega _{\nu} $ plane for $H_0 = 50 $.
First, we should note that no CHDM model fits the Bough \& Efstathiou
data too well (except for $H_0=40$ or low $n$ ). In particular,
$\chi ^2 < 2 $ is achieved only for $H_0 \le 50 $ if the spectral
index is $n=0.95$. Second, $\chi ^2 $ contours for these two
tests have a rather close structural resemblance. Since the number of
$P(k)$ points used is small and the errors
given by Bough \& Efstathiou (1993) may be questioned, we don't
use this test to set specific restrictions on the model.
However, using a rather relaxed limitation $\chi^2 < 4 $ for
the angular correlation function test, we can confirm if not strengthen
the lower limit on $\Omega_{\nu}$  previously obtained from the
counts-in-cells data for parameters in the most interesting
region around $H_0=50,~\Omega_{\nu}=0.25 $.
Other conclusions made as the deterioration of
the fit with increasing of $n$ or $H_0$ are also confirmed by this test.

In the previous consideration, we have not discussed the size of the
allowed region in the amplitude $A$ dimension of the parameter space.
In Fig. 4, we present the $\Omega_{\nu} - A$ cross-section of the
parameter space with fixed $H_0=50$.
The shaded strip selects the adiabatic amplitude $A$
following from the COBE result $(\Delta T/T)_4 = (12.8 \pm 2.3) \,
\mu {\mathrm K}/T_{\gamma}$.
Now it depends on $n$ (according to Eq. (\ref{dt}))
but not, practically, on $\Omega_{\nu}$. This is also the case for
the estimation of bulk velocities $v(R)$ in spheres of
radii $R=80,~120 h_{50}^{-1} \, {\mathrm Mpc} $.
The condition $\sigma_8<0.67 $ and the quasar number test (\ref{s11})
for $M=10^{11} \, {\mathrm M}_{\odot} $ leave the triangular area left and
below the point of intersection of solid curves as the allowed range
of parametrs. Exactly this point of intersection produces the upper dashed
line in Figs. 1,~2.

In Fig. 4, we see new aspects of the failure of the CHDM model to be
successful for large $n$. Not only the $\sigma_8$ condition can't be
fulfilled, but also large-scale bulk velocities become
simultaneously too high to be compatible with
$\sigma_8<0.67$ for $n\ge 1.1$ and too low
in comparison with the COBE amplitude for $ n > 1.2$.

We understand that there are two effects which might make these limits
less strong. First, some part of the observed $\Delta T/T$ can be due
to gravitational waves. Then the adiabatic amplitude $A$
derived from
the {\it COBE} data will be lower than that in Fig. 4.
Second, bulk velocities in our vicinity can differ from average
ones in the Universe by cosmic variance.
Therefore, assuming {\it both} that the velocities $v(R)$ given by
POTENT are at least 40\% larger than their {\it rms} values
and that the primordial gravitational wave background is responsible
for the increase of the observed large-angle $\Delta T/T $ by at least
$1/3$, one can, in principle,
reconcile the amplitude $A$ from these tests with $\sigma_8<0.67 $ for
$n=1.2$. However, the former assumption is equivalent to introducing
one more fundamental dimensionless parameter that shifts the model to
the fourth
level according to our classification. Really, since adiabatic and
tensor fluctuations are statistically independent, we have
$\langle(\Delta T/T)^2\rangle_{tot}  = \langle(\Delta T/T) ^2
\rangle_{\mathrm ad} + \langle(\Delta T/T )^2\rangle_{\mathrm gw}$.
Therefore, to increase the {\it rms } value of $(\Delta T/T)_{tot}$
by $40\%$ or more, $\langle(\Delta T/T )^2 \rangle_{\mathrm gw}$ should
be no less than $\langle(\Delta T/T) ^2 \rangle_{\mathrm ad}$ - a kind
of additional ``fine tuning'' in the $n>1$ case (in contrast to
the $n<1$ case where such a condition arises naturally).
According to the main idea of classification of
cosmological models presented in the Introduction, this does not mean that
such a model is bad (because dimensionless fundamental parameters
may be fine-tuned to some number), it simply shows that more
new significant fundamental parameters are necessary for explanation of
observational data than it was assumed initially. Thus, this way
is not admissible if we want to stay among models of the third level.

This can be illustrated using the simple inflationary model presented
in the end of Introduction. Under the condition $|\phi|\ll M_P$
assumed earlier, the gravitational wave contribution is small:
assuming $n\approx 1$ and using well-known expressions
for the slow-roll motion, it is straightforward to show that
\begin{equation}
{T\over S}\equiv {\langle ({\Delta T\over T})_{lm}^2\rangle_{\mathrm gw}
\over \langle ({\Delta T\over T})_{lm}^2\rangle_{\mathrm ad}}=
\left(4.16{m^2\phi\over H_1^2M_P}\right)^2\left(1+{m^2\phi^2\over
2V_0}\right)^{-2}
\end{equation}
for $2\ll l\le 30$ (the numerical coefficient in round brackets
is $6\%$ more for $l=2$, but $6\%$ less for $l=3$ and $9\%$ less
for $l=5,~6$). Here the
value of $\phi$ is, as usually, taken at the moment of the first
horizon crossing during the inflationary stage ($\approx 60$ e-folds
before its end). Now let us try to get a larger value of $T/S$ by
assuming that $|\phi|\sim M_P$ at this moment. To achieve this,
we have to assume some specific relation between $\phi_f$ and $M_P$.
Earlier, a value of the dimensionless constant $\phi_f/M_P$
was not important for comparison with observational data, it was
enough to assume that it was sufficiently small. But now this value
becomes the new significant fundamental parameter of the model.
Of course, then the power spectrum cannot
be considered as an exactly power-law one. Note that,
incidentally, this specific model of the fourth level does
not achieve the aim of having $T>S$ for $n>1.1$, too,
if we assume the abovementioned upper limit on the spectral index
at very small scales following from the absence of excessive PBH
formation: $n\le 1.4$. This requires $m\le 0.8H_1$, so, for scales
where the local index $n>1.1$, the temperature
fluctuation amplitude increases by $\sqrt {1+{T\over S}}<1.2$.

For values of the spectral index $n<1$, the situation with bulk
velocities becomes inverse. For $n \le 0.85$,
they are higher than those following from the normalization
to the large-scale $\Delta T/T$. Now the possible
effect of gravitational waves only worsens the discrepancy
(for $n=0.85$, in the framework of inflationary models with
the exponential inflaton potential, the adiabatic amplitude $A$ is
$40\%$ smaller than the one given in Fig. 4 due to
this effect). Therefore, following Liddle \& Lyth (1993),
we exclude models with $ n < 0.9 $.

\section{Conclusions and discussion}

We compared the CHDM model with a falling power-law primordial
spectrum of adiabatic perturbations ($n>1$) with observational data.
This model has one more adjustable dimensionless parameter ($n-1$,
which can be expressed in terms of an additional parameter of an
inflaton potential) than the CHDM model with the approximately flat
($n\approx 1$) spectrum. It might be thought naively that the model
with more parameters fits data better. Remarkably, we found just the
opposite: fit to the data becomes {\em worse} with the growth of
$n-1$, and may be considered as unreasonable already for $n>1.1$.

Combining this with the previously known result that the CHDM model is
hardly compatible with observations for $n< 0.9$ (Liddle \& Lyth 1993,
PS), we arrive to the conclusion that the CHDM model requires
the approximately flat ($|n-1|<0.1$) primordial spectrum among
all possible power-law spectra of adiabatic perturbations. This shows
the robustness of the CHDM model with the simplest inflationary
initial conditions. Our conclusion agrees qualitatively with that
in the recent paper by Lyth \& Liddle (1994). Moreover, the best fit
to the data is achieved for $n$ slightly less than $1$, and around
values of the effective slope expected in the simplest inflationary
models: either with a scalar field with a polynomial
potential $V={m^2\phi^2\over
2},~n\approx 0.97$ or $V={\lambda \phi^4\over 4},~
n\approx 0.95$ (chaotic inflation, Linde 1983), or in the
higher-derivative gravity
$R+R^2$ model (Starobinsky 1980) where $n\approx 0.97$, too, or
in the new inflationary model with the Coleman-Weinberg potential
(Linde 1982, Albrecht \& Steinhardt 1982) with $n\approx 0.95$.

Vice versa, the CHDM model presents the best possibility for the
realization of these inflationary models, because the other
alternative, the CDM+$\Lambda$ model, has more serious problems.
The most promising purely CDM models
seem now to be based on a non-scale-invariant, step-like
initial spectrum of adiabatic perturbations produced in
more complicated inflationary models (Gootl\"ober, M\"ucket
\& Starobinsky 1993, Peter, Polarski \& Starobinsky 1994), they
belong to the third level in our classification.
Thus, unexpectedly, the fate of the simplest inflationary
models appears to be closely tied to the fate of the CHDM model.

In addition, there is a place in the CHDM model with $0.9<n<1$
for a modest but noticeable gravitational wave contribution to
large-angle $\Delta T\over T$ fluctuations expected for
chaotic inflationary models with polynomial inflaton potentials
(but not for the new inflationary model and the $R+R^2$ model).
It leads to the increase in the total {\it rms} amplitude
of these fluctuations by, e.g., $\sqrt{1+{T\over S}}-1
\approx 10\%$ for the $V=
{\lambda \phi^4\over 4}$ inflationary model (Starobinsky 1985).
Note that all the simplest inflationary models listed above
do not require additional fundamental parameters to specify
their predictions for (weakly scale dependent) values of
the spectral index $n-1$ and the ratio $T/S$. So, they belong
to the second level if combined with the CHDM model for dark matter.
The increase in $\Delta T\over T$ may be even found desirable in view
of recent papers by
G\'orski et al. (1994), Banday et al. (1994) where further rise
of the COBE results up to $Q_{rms-PS}=(19-20)~\mu {\mathrm K}$
is suggested. On the other hand, it is clear that
a significanly larger gravitational wave contribution is incompatible
with the CHDM model.

Of course, the most crucial confirmation of the CHDM model would
be a direct or indirect (through neutrino oscillations parameters)
discovery that the $\tau$-neutrino mass is really around $5$ eV as
predicted by the best fit to the model. But of no less importance
is the precise determination of the Hubble constant because the
model can't work for $H_0>60$ km/s/Mpc, and it is better to have
its value around $50$ km/s/Mpc. The third critical test
for the model is the abundance of galaxies and quasars at large
redshifts.

Finally, let us return to the case of two (or even three)
comparable neutrino
masses mentioned in the Introduction. If neutrino concentrations
are the standard ones and $\Omega_{\nu}$ denotes the total
neutrino energy density in terms of the critical one, then what
stands in the left-hand side of Eq. (\ref{mass}) is actually
the {\em sum} of masses of all neutrino species. In this case,
the relative
transfer function $D(k,\Omega_{\nu},z)$ (\ref{transf}) will
have the same step-like form discussed in PS, but the transition
region from one plateau to another will be shifted to larger scales
due to increase of $R_{nr}$. So, in the first approximation, we
may expect that the best fit is given by the condition that the
sum of all neutrino masses should be about $5$ eV. However, due
to the abovementioned shift of characteristic scales in the transfer
function, more caferul analysis of this higher level model is needed
that will be carried elsewhere.

{\bf Aknowledgments}

A.S. is grateful to Profs. Y. Nagaoka and J. Yokoyama for their
hospitality
at the Yukawa Institute for Theoretical Physics, Kyoto University.
A.S. was supported in part by the Russian Foundation for
Basic Research, Project Code 93-02-3631, and by Russian Research
Project ``Cosmomicrophysics''. D.P. is grateful to Profs. Alan Omont
and Francois Bouchet for their hospitality
at the Institute d'Astrophysique de Paris where this
project was started as well as to Prof. Simon White and Dr. Thomas Buchert
from Max-Planck-Institut f\"ur Astrophysik, M\"unchen where the
final version of the paper was prepared.

\clearpage

{\bf Figure captions}

\begin{description}
\item[Fig.1]
Restrictions in the $ H_0 - \Omega_{\nu} $ parameter plane
for $n=0.85,~0.95$ (upper row) and $n=1.1,~1.2 $ (lower row)
following from:
a) Stromlo-APM counts-in-cells test. Solid lines correspond to
$\chi ^2 = 2,~7 $ contours.
b) The $\sigma_8$ condition.
The values $ \sigma_8<1,~0.67$ are achieved left to the dashed
(correspondingly right and left) lines.
c) The combination of the $\sigma_8<0.67$ condition with
quasar and galaxy formation conditions (dotted lines).
The allowed region lies below the upper dotted line
if the limitation set for objects of the mass
$ M = 10^{11} \, {\mathrm M}_{\odot} $ is used
and below the lower one if $ M = 10^{12} \, {\mathrm M}_{\odot} $.
\item[Fig.2]
Same tests as in Fig. 1 are given in the $ n - \Omega $ parameter
plane for $H_0=40,~50 $ (upper row) and $H_0=60,~70 $ (lower row).
\item[Fig.3]
Comparison of the $\chi^2$ contours for fits to the
cloud-in-cells values (solid contours $\chi ^2 = 2,~7$)
and the power spectrum $P(k)$ from the galaxy angular correlation
function (dashed contours $\chi ^2 = 1,~2,~4$).
The left panel shows the $ H_0 - \Omega _{\nu} $
plane for $n=0.95$,
the right one shows the $n - \Omega _{\nu} $ plane for $H_0=50$.
\item[Fig.4]
Amplitude of perturbations $ A $ in the CHDM model
as follows from {\sc potent} data for bulk peculiar velocity $ v(R) $
[regions between dotted and dashed lines correspond to
$\pm 1 \sigma $ error bars for $ v(80 h_{50}^{-1} \, {\mathrm Mpc}) $ and
$ v(120 h_{50}^{-1} \, {\mathrm Mpc}) $ respectively]
compared to the {\it COBE} limits [shaded region].
The region {\it above} the solid line (\romannumeral1) is allowed by
the quasar
number condition $ f( \ge 10^{11} \, {\mathrm M}_{\odot},z=4 ) \ge 10^{-4} $.
The region {\it below} the solid line (\romannumeral2) is allowed by the
condition $\sigma_8<0.67$.
\end{description}
\end{document}